\documentclass[10pt,twocolumn]{article}   

\usepackage[a4paper]{geometry}
\usepackage{graphicx}

\usepackage{epstopdf} 
\usepackage{epsfig}
\usepackage{color}
\usepackage[compress,sort]{cite}

\begin{document}
\onecolumn
\begin{center}
\noindent
\textbf{\large{XMCD studies of thin Co films on BaTiO$_3$}}\\
\end{center}

\noindent
M Welke$^1$\footnote{martin.welke@uni-leipzig.de}, J Gr\"afe$^1$\footnote{Present address:
Max-Planck-Institut f\"ur Intelligente Systeme Stuttgart, Heisenbergstra{\ss}e 3, 70569 Stuttgart},
R~K~Govind$^2$, V~H~Babu$^1$\footnote{Present address:
CELLS - ALBA Synchrotron, Carretera BP 1413, de Cerdanyola del Valles a Sant Cugat del Valles, 08290 Cerdanyola del Valles, Barcelona}, M~Trautmann$^2$, K-M~Schindler$^2$ and R~Denecke$^1$\footnote{denecke@uni-leipzig.de}
\vspace{10pt}

\noindent
{$^1$~Universit\"at Leipzig, Institut f\"ur Physikalische und Theoretische Chemie, Linn\'estra{\ss}e~2, 04103 Leipzig, Germany}\\
{$^2$Martin-Luther-Universit\"at Halle-Wittenberg, Institut f\"ur Physik, Von-Danckelmann-Platz 3, 06120 Halle, Germany}

\vspace{10pt}

\noindent

This is an author-created, un-copyedited version of an article accepted for publication in J. Phys.: Condens. Matter. IOP Publishing Ltd is not responsible for any errors or omissions in this version of the manuscript or any version derived from it. The Version of Record is available online at:\newline 
http://dx.doi.org/10.1088/0953-8984/27/32/326001

\vspace{10pt}

\begin{abstract}
Different layer thicknesses of Cobalt ranging from 2.6~\r{A}~(1.5~ML) up to 55~\r{A}~(30.5~ML) deposited on ferroelectric BaTiO$_3$ have been studied regarding their magnetic behavior. The layers have been characterized using XMCD spectroscopy at remanent magnetization. After careful data analysis the magnetic moments of the Cobalt could be determined using the sum rule formalism. There is a sudden and abrupt onset in magnetism starting at thicknesses of 9~\r{A}~(5~ML) of Cobalt for measurements at 120~K and of 10~\r{A}~(5.5~ML) if measured at room temperature. Initial island growth and subsequent coalescence of Co on BaTiO$_3$ is suggested to explain the sudden onset. In that context, no magnetically dead layers are observed. 
\end{abstract}

%\noindent{\it Keywords\/}: \\

\newpage
\twocolumn
\section{Introduction}

Systems with multiferroic properties are of increasing interest for research. \cite{ramspal} In the beginning of this research field single phase materials exhibiting several ferroic ordering phenomena were studied. Theory using first-principle calculations has shown the possibility of simultaneous switching of different ferroic properties by a strictly electronic mechanism. \cite{duan} For example, in bismuth ferrite (BiFeO$_3$) or yttrium manganite (YMnO$_3$) magnetic switching of their polarization can be achieved and vice versa at low temperatures. \cite{chemo, wang1} Materials with multiferroic properties would open the door to numerous applications, e.g. phase transformers, microwave filters, or RAM components. \cite{ramspal, chemo, wang1, wang2}

However, single phase materials with ferromagnetic and ferroelectric properties are scarcely to be found. This results from the origin of these two effects. In common cases, ferroelectricity can be found in transition metal compounds and requires formally empty \textit{d} orbitals. In those compounds a common theory states that the ferroelectric effect is created by displacement polarization. On the other hand ferromagnetism is mainly dependent on ordering of the spins within the \textit{d} orbitals of a transition metal requiring them to be partially filled. \cite{hill2}

Considering the origin of ferromagnetism and ferroelectricity and the subsequent limited number of single phase multiferroics, the field of research shifted towards layer systems. In these systems each layer is either ferromagnetic or ferroelectric. \cite{ramspal} An additional benefit of such systems would be that across the interface the interaction between ferroelectric polarization and magnetic polarization could be investigated systematically.

In this wide field of research various layer systems are investigated. Among them are studies using barium titanate (BaTiO$_3$), which is ferroelectric below 393~K, as  substrate for deposition of various ferromagnetic or ferrimagnetic materials. \cite{sterbi, brivio1, bocher, graefi} Additionally, thin magnetic films have been reviewed several times while different substrates have been used. \cite{vaz2, lawes, wang2} In a first step the magnetic moments of materials deposited as thin layers on those substrates are of interest. When a magnetic moment can be found detailed studies towards a coupling need to be done. Recent studies of Fe on BaTiO$_3$(001) showed promising results towards such layer systems. \cite{brivio1, bocher} In the long term, purely oxidic systems, e.g. magnetic ferrites such as CoFe$_2$O$_4$, are of special interest due to their more robust nature and existing studies on three-dimensional structures. \cite{ramspal}

Different ferromagnetic materials are tested on BaTiO$_3$ in order to have a bigger variety. Therefore, in this work thin layers of Cobalt have been studied regarding their magnetic moment. Cobalt differs from well-studied Iron \cite{bocher, brivio1, govind2, govind1, lahti, shira} regarding its crystal structure, lattice parameters, and magnetic moments. However, it is very similar to Iron regarding its chemical properties and can, hence, be used as comparative classical ferromagnetic system. In Fe, there is a clear onset in magnetism which would be expected for Co as well. \cite{govind1} Additionally, Cobalt has two modifications being hexagonal at room temperature and cubic above 700~K. \cite{yoo} %\textcolor{red}{The results of both pure metallic layers might help to further understand the spectra of magnetic ferrites with mixed elements}, e.g. CoFe$_2$O$_4$ and should be understood as starting point for a broader field of research. \textcolor{red}{These oxide spectra should exhibit different shapes showing their oxidic nature.}

Future applications will need to fulfill a good and stable remanent magnetization at low cost and material input (i.e. the smallest dimensions). Additionally, spectroscopic methods are limited in information depth, which requires (ultra)thin layers so that the influence of the interface can still be investigated. Furthermore, that influence might become smaller, when the overlayer gets thicker. Hence, a range of layer thicknesses has been chosen for this work to cover because of a proposed thickness-dependent onset in magnetism within thin magnetic layers. Measurements included a wedge of up to 12.6~\r{A}~(7~ML) as well as thicker Cobalt films of 12.5, 15, 17.5, 21.5 and 30.5~ML. X-ray magnetic circular dichroism (XMCD) was used to investigate the magnetic properties of the resulting layered systems. Calculation of the magnetic moments from the XMCD data was performed using the sum rule formalism. \cite{carra} Here, the Cobalt~\textit{L} edge was used.

\section{Methods}
Barium titanate single crystals with a nominal surface orientation of (001) used as substrate were sputtered with argon (2$\times$10$^{-5}$~mbar with a sample current of 2.5~$\mu$A) for a period of 20~minutes. Afterwards the samples were annealed at 1100~K in a 1$\times$10$^{-6}$~mbar atmosphere of oxygen for 10~minutes. Using low energy electron diffraction~(LEED) the cleanliness of the crystal surfaces was checked. There are clear reflexes of the cleaned sample. This information is supported by XAS measurements at the different absorption edges, which resemble spectra published in literature very well. \cite{sori} Although the surface termination has not been determined experimentally for the current sample, it is quite likely that a BaO termination was achieved by the preparation method, as was reported in different publications. \cite{berlich, pancotti} It also need to be mentioned that although the substrate surface was nominally (001), due to the spontaneous polarization the surface usually contains both a and c domains. This is expected to influence the growth of Co. Since no very good match of the lattice constants is given anyway, this influence is considered here to be of minor importance.

A SPECS EBE-4 electron beam evaporator was used to evaporate a Cobalt wedge from 1.5~ML~(2.6~\r{A}) up to 7~ML~(12.6~\r{A}) with a linear slope over a range of around 8~mm as well as different layers produced individually (12.5, 15, 17.5, 21.5 and 30.5~ML) onto the barium titanate at a base pressure of 5$\times$10$^{-9}$ mbar, with subsequent annealing at 430~K for each layer. The setup used yielded a growth rate of 0.7~\r{A}/min. These growth rates have been determined using a quartz crystal micro balance.

X-ray absorption spectroscopy (XAS) measurements in total electron yield (TEY) were carried out at room temperature at synchrotron sources BESSY~II, Berlin at beamline UE56/2-PGM-2 and MAXlab~II, Lund at beamline I1011. At room temperature barium titanate has a tetragonal crystal lattice with lattice parameters a~=~b~=~3.991~\r{A} and c~=~4.035~\r{A} while being ferroelectric. Additionally, at room temperature bulk Cobalt has a hexagonal configuration with lattice parameters of a~=~b~=~2.514~\r{A} and c~=~4.105~\r{A}. Although there could be a small averaging of layers at the wedge due to the spot size of the beam it should only effect a difference of $\pm$~1~ML of Co. Taking the beam diameter of less than 100 $\mu$m and the shallow wedge the probability of averaging between two adjacent integer layer thicknesses is very low. Cobalt was magnetized \textit{in-plane} using an electromagnet supplying a magnetic field of 50~mT at BESSY or 70~mT at MAXlab, respectively, for measuring the XMCD at remanence. All the layers were measured at room temperature (300~K) as well as 120~K. In order to measure the XMCD signal, the helicity from the undulators was changed in successive spectra. After normalization to the photon flux the difference spectra can be calculated. Details will be described below.

\section{Results}

\begin{figure}[!htb]
\begin{center}
\includegraphics[width=8cm]{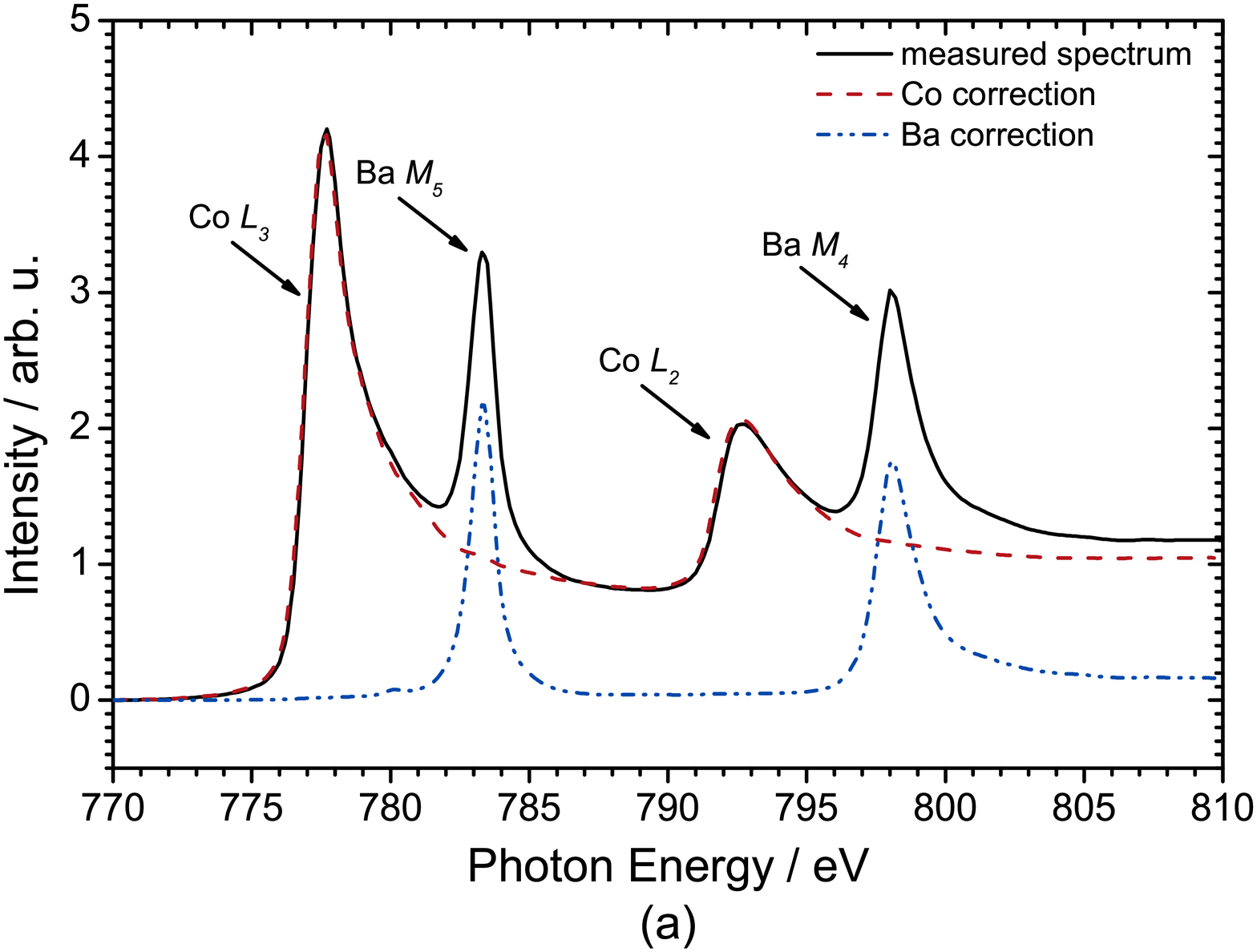}
\includegraphics[width=8cm]{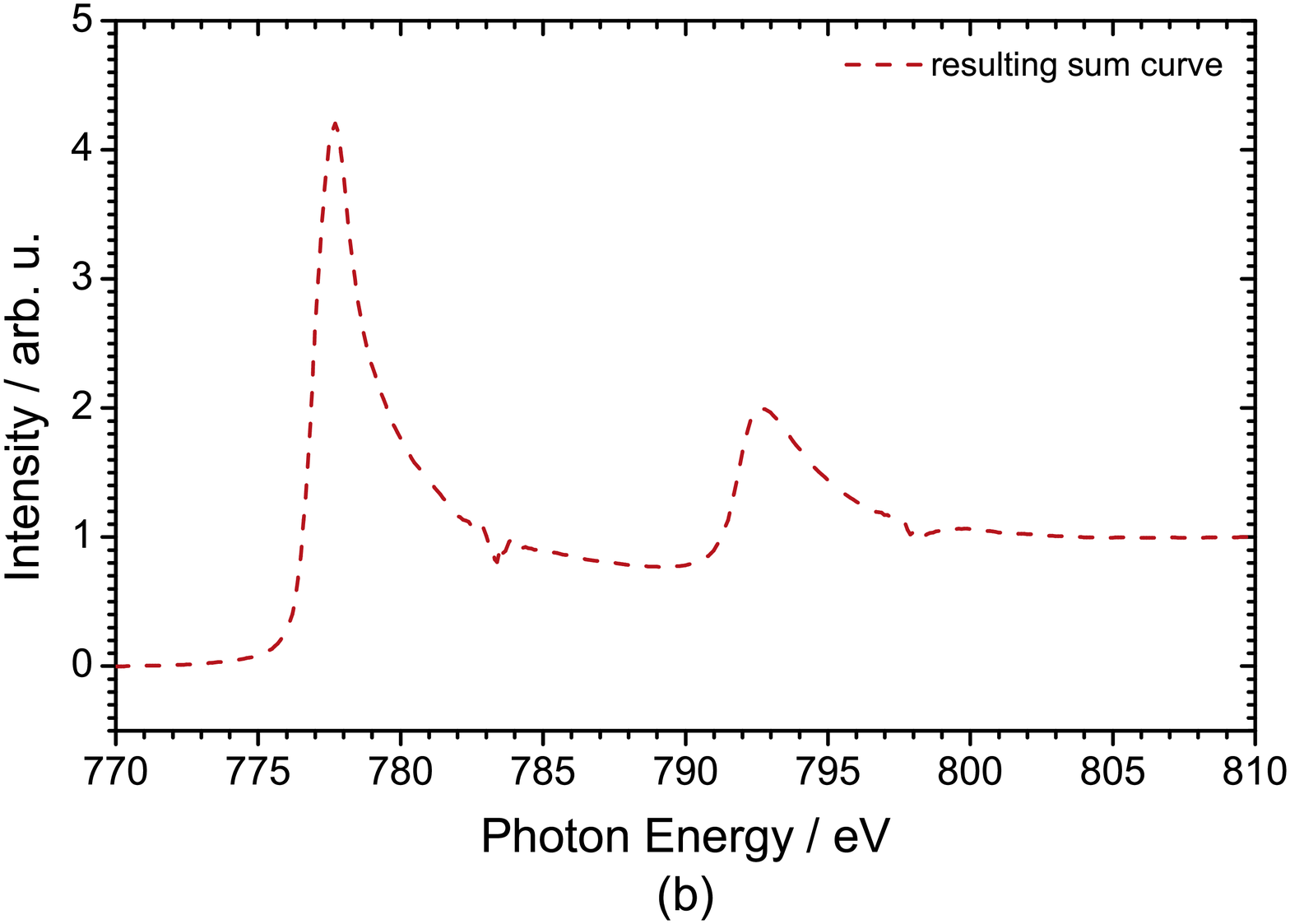}    		 				  
\caption{Linear combination of pure Co~\textit{L}~edge spectrum of a thick Co layer (red dashed line in (a)) and pure Ba~\textit{M}~edge spectrum in barium titanate (blue dash-dotted line in (a)). In (a) both of the pure spectra as well as the measured spectrum (black solid line) are shown. In (b) the resulting Ba free sum curve of the Co~\textit{L}~edge after subtraction of the Ba~\textit{M}~edge portion is provided.}  
\label{fig1}                                 					
\end{center}                                 					
\end{figure}

After deposition, the chemical identity of the films was verified using XA spectra at the Cobalt~\textit{L}~edge, the Barium~\textit{M}~edge, the Titanium~\textit{L}~edge, and the Oxygen~\textit{K}~edge. Figure~\ref{fig1} shows a typical spectrum containing both the Co~\textit{L}~and Ba~\textit{M}~edges. In these spectra there is a separation of 5.8~eV between the Co~\textit{L}~edge and the Ba~\textit{M}~edge, leading to a substantial overlap at both, Co~\textit{L$_3$}~/~Ba~\textit{M$_5$} and Co~\textit{L$_2$}~/~Ba~\textit{M$_4$}. For further analysis a proper algorithm to separate the contributions had to be developed, as will be described here. Typical shoulders expected for oxidized species could not be observed in the Co spectra (see lower photon energies of the Co~\textit{L$_3$}~edge at figure~\ref{fig1}(a)). \cite{magnu}

The use of X-ray photoelectron spectroscopy (XPS) does not improve the overlap situation. In XPS the variation in binding energy between Co~2\textit{p} and Ba~3\textit{d} was measured with only 0.4~eV and, thus, much smaller than for metallic Co and Ba. \cite{hudson2, lebugle} Consequently, given the intrinsic widths of the peaks it is not possible to evaluate XP spectra where the Cobalt layer thickness is below the saturation thickness.

As pointed out before Cobalt has a hexagonal crystal lattice at room temperature. On the other hand, barium titanate is in a tetragonal configuration so that epitaxial growth would be difficult. Since no LEED pattern was obtained for neither the wedge-shaped layer nor the individually produced layers the structure of the film cannot be determined. Here, it can be assumed that small islands of hexagonal configuration are growing on top of barium titanate. With increasing nominal layer thickness these islands are supposed to coalesce.

\begin{figure}[!htb]
\begin{center}
\includegraphics[width=8cm]{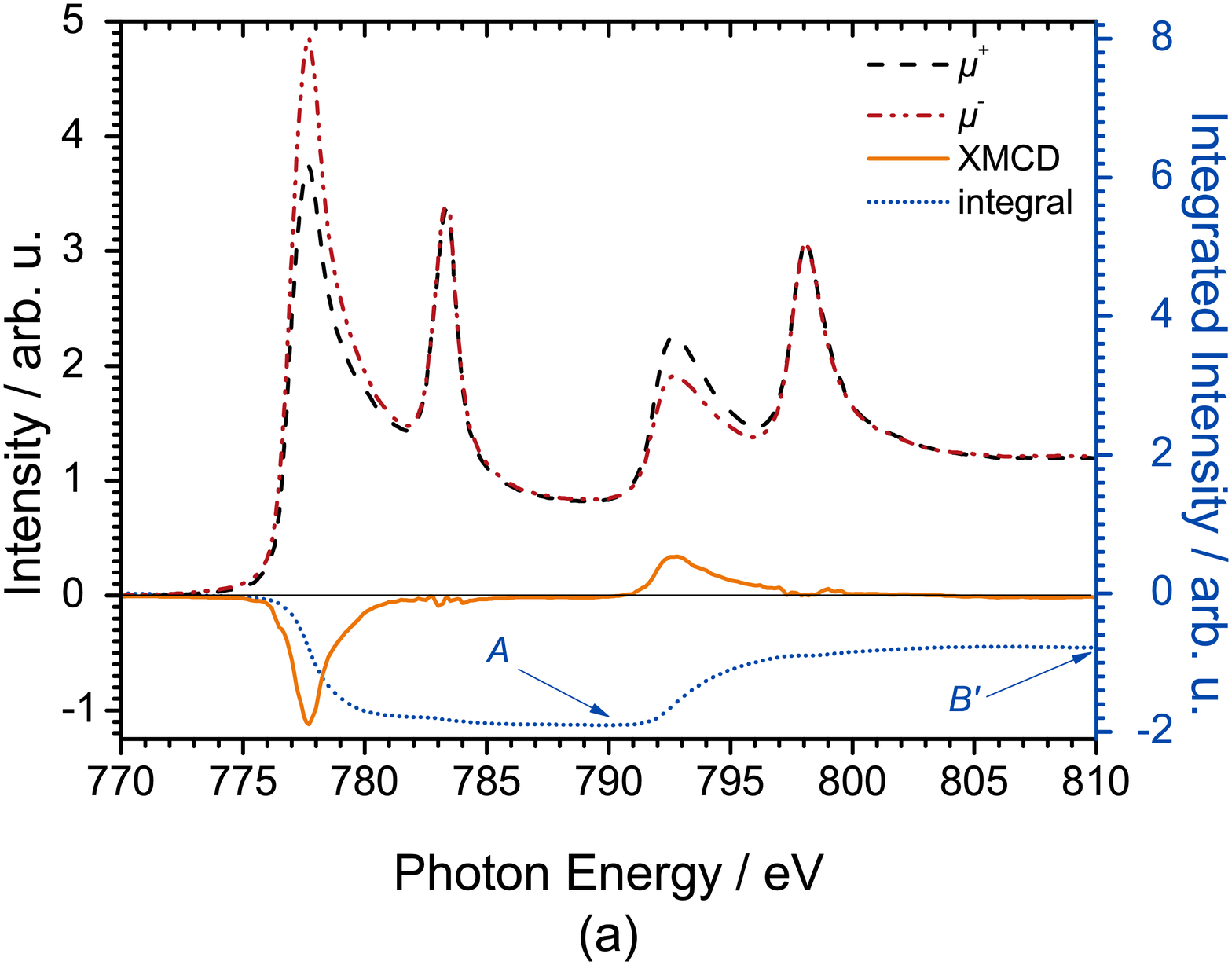}
\includegraphics[width=8cm]{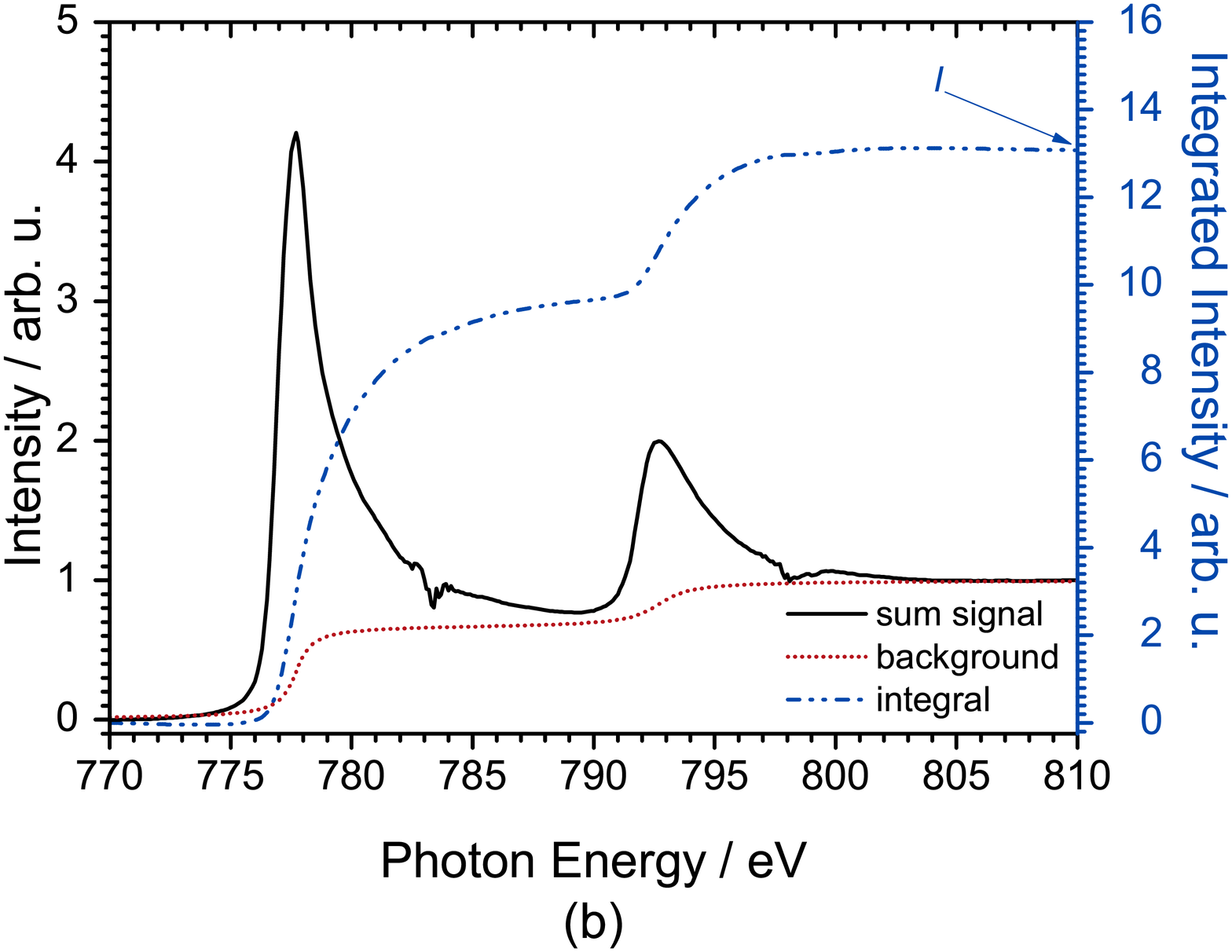}    		 				  
\caption{The parameters for the sum rule analysis are taken from the respective graphs (see text). In (a) the parameters $A$ and $B'$ are taken from the integrated difference signal (blue dotted line) while it is shown that the Barium~\textit{M}~edge does not have any influence on that difference signal (orange solid line) or its integral. In (b) the parameter $I$ provided by the integral (blue dash-dotted line) of the created white line (sum function (black solid line) minus background (red dotted line)) is provided.}  
\label{fig2}                                 					
\end{center}                                 					
\end{figure}

A direct application of the sum rules according to \textsc{St\"ohr} \cite{IBMsto, jelcsto, jmmmsto} on the measured XAS spectra is not possible due to the presence of the Ba~\textit{M}~edge. A pure Co spectrum has been extracted by using a linear combination of a Co~\textit{L}~edge reference and a clean Ba~\textit{M}~edge reference. In figure~\ref{fig1}~(a) the steps and the resulting separate spectra of such a linear combination are shown for the example of a Cobalt sum curve. These references were taken from cleaned barium titanate (blue dash-dotted Ba line) and a thick Cobalt layer (22~ML, red dashed line), respectively. In figure 1(b) the spectrum obtained by subtracting the resulting Ba contribtion from the measured spectrum is shown. The resulting Cobalt spectrum (see in figure~\ref{fig2}~(b)) could be used for further evaluation.

\begin{figure}[!htb]
\begin{center}
\includegraphics[width=8cm]{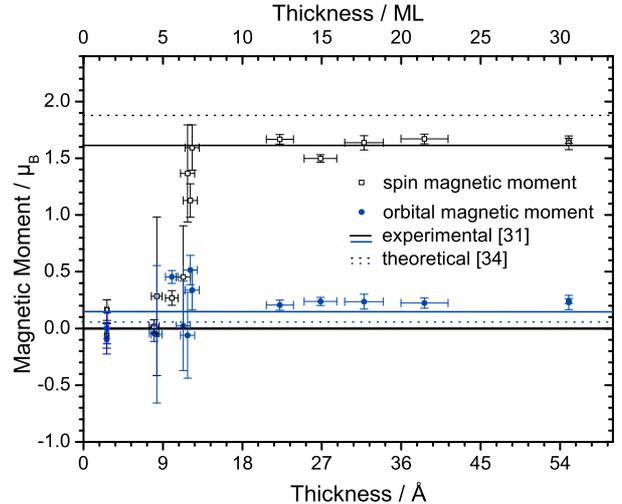}   		 				  
\caption{Thickness dependence of the spin magnetic moment (black open squares) and the orbital magnetic moment (blue closed circles) measured at 300~K. Both moments are increased compared to known experimental bulk values (black and blue solid lines).~\cite{chen1} Compared to a calculated relaxed, free-standing Cobalt monolayer (black and blue dotted lines) the orbital magnetic moment is increased whereas the spin magnetic moment is decreased.~\cite{lehnert}}
\label{fig3}                                 					
\end{center}                                 					
\end{figure}

Furthermore, that linear combination has been used as an internal comparison of the Cobalt thickness. A quartz crystal micro balance was used to calibrate the growth rate. Additionally, that calibration has been cross checked by a \textsc{Lambert-Beer} approch using respective damping factors for the electrons from literature and was shown to be correct within $\pm$~0.3~\r{A}. \cite{eal, imfp} Different uniform layer thicknesses have been prepared first. With these layers a model has been applied providing the nominal layer thickness of Cobalt as a function of the linear combination factor. The layer thicknesses derived from the linear combination factor are used for further description of the data. Since there is still some statistical error within that routine the absolute value might differ, which is represented in the error bars seen in figures~\ref{fig3}~and~\ref{fig5}. Nevertheless, the thick layers exhibit similar total magnetic moments and can, thus, be compared to each other. The layer thicknesses calculated by this routine reach from 2.6~\r{A}~(1.5~ML) up to 55.0~\r{A}~(30.5~ML) of Cobalt (using the nominal value from the quartz crystal micro balance if no Ba signal could be detected anymore due to absorption within the Co layer). One monolayer of Cobalt was taken as 1.8~\r{A} for calibrating monolayer~(ML) thicknesses.

Using the corrected Co~\textit{L}~edge data the magnetic properties of Cobalt can now be determined using XMCD spectroscopy and sum rule analysis. Therefore, careful data processing has to be applied, because both difference and sum spectra are needed. \cite{carra} In order to obtain the sum signal from the measurements with negative polarized and positive polarized light, the two spectra are added and divided by two. Afterwards the created spectrum is corrected for Ba as pointed out above. Additionally, the sum spectrum is normalized to a total step height of 1 after subtraction of a common constant pre-edge background for both helicities. Both single spectra have to be corrected in the same way. However, when this sequence is applied there is an offset at the post-edge of the single spectra. This behavior originates in the remanent magnetic moment of the cobalt and the resulting magnetic field or small instabilities at the focusing mirrors and differences in the helicities. Electrons emitted from the sample can be driven away from the surface by the stray field of the remanent magnetization or even pulled back to the surface. Therefore, \textsc{Goering} \textit{et al.} suggest a mathematical solution which was applied to the data and which will be briefly outlined here. \cite{goer1}

The measured TEY signal $I^\pm$ is proportional to the x-ray absorption coefficient $\mu^\pm$, which is composed of a nonmagnetic absorption $\mu_0$ and a magnetization dependent absorption $\mu_c$. Electrons leaving the sample will be affected by the stray field of the remanent magnetization. For one helicity, the electrons are pulled back into the sample whereas in the other case they are pushed out of it. In order to correct this effect the post-edge detection becomes important. At photon energies around 10~eV away from the absorption edge the magnetization dependent contribution is assumed to be approaching zero. In the end, the ratio of $I^{+}/I^{-}$ at this position provides a correction factor ($\Delta f$) applied to $I^{+}$ for magnetic field correction converging both spectra in the post-edge region. However, a very small portion of the XA spectrum will then be transferred into the difference signal although this is not visible in the end.

In figure~\ref{fig2} an example of the result of the data processing is shown. In case of the difference signal (orange solid line in figure~\ref{fig2} (a)) the linear combination was not necessary as the Ba~\textit{M}~edge cancelled out. As a result, that procedure was applied to the sum spectra only as there are no changes found for the difference signal. After performing the linear combination and subtraction of the Ba~\textit{M}~edge the integral of the white line where the step background is subtracted from the sum curve (blue dashed line in figure~\ref{fig2}~(b)) as well as the integral of the difference curve (orange solid in figure~\ref{fig1}~(b)) were generated. 

These were then used in the sum rule formalism. \cite{IBMsto, jelcsto, jmmmsto} Furthermore, the difference signal was corrected for the grazing angle as well as degree of polarization of the light in order to consider the effective magnetization parallel to the light incidence direction and non-unity degrees of polarization where needed, respectively. \cite{chen1} In fact, the grazing angle was in between 30$^{\circ}$ and 50$^{\circ}$, while the polarization of light was in between 90~$\%$ and 95~$\%$ depending on the beamline used.

Using parameters derived from the analysis of the spectra, the spin ($m_S$) and orbital ($m_O$) magnetic moments can be calculated in units of $\mu_B$ as given in equations~(\ref{e1})\hbox{-}(\ref{e4}).
\begin{eqnarray} \label{e1}
{{m_S}} = -{{[A - 2B]}\over {C}}{\mu_B}
\end{eqnarray}
\begin{eqnarray} \label{e2}
{{m_O}} = -{2{[A - B]}\over {3C}}{\mu_B}
\end{eqnarray}
with
\begin{eqnarray} \label{e3}
{{C}} = {{I}\over {n_d}}
\end{eqnarray}
and
\begin{eqnarray} \label{e4}
{{B}} = {{B' - A}}
\end{eqnarray}
The parameters $A$ (at 790~eV from integrated difference), $B'$ (at 810~eV from integrated difference), and $I$ (at 810~eV from integrated white-line) were taken from the spectra as shown in figure~\ref{fig2}. The number of \textit{d}~holes for Co needed for calculation of the magnetic moments was estimated to be $n_d$~=~2.49, which has been calculated for bulk Cobalt. \cite{chen1, guo} Errors within the number of \textit{d}~holes originating in the layer thickness are comparably small compared to the errors created by statistical noise of the spectra. Therefore, they are well included in the error bars for the resulting magnetic moments.

Following the careful data processing, the normalized XMCD spectra are directly comparable to each other. A sudden onset of the XMCD at around 10~\r{A} is observed for room temperature measurements, which is at thicker layers than observed for Co on oxides before. \cite{sass} Below that thickness there is no significant effect whereas above 12.6~\r{A} all the normalized XMCD spectra have approximately the same intensity. In figure~\ref{fig3} the thickness dependence of the spin magnetic moment and the orbital magnetic moment is shown providing a clear onset in magnetism. At around 10~\r{A}~($\approx$~5.5~ML) the first reliable total magnetic moments are found. Furthermore, a bulk like magnetic moment has evolved at 12.6~\r{A}~(7~ML). In figure~\ref{fig4} we show the thickness-dependent difference spectra for three characteristic thicknesses, from which the calculated magnetic moments are included in figure~\ref{fig3}. At 8~\r{A} there is still no observable XMCD whereas at 10~\r{A} it can be seen easily. The full XMCD occurs at 12.6~\r{A}. An example of the corresponding full XMCD signal is shown at a layer thickness of 22~\r{A} in figure~\ref{fig4}. Because of the statistical noise the data points in figure~\ref{fig4} have been smoothed using a \textsc{Cubic B-Spline} function in OriginPro 8 as a guide to the eye. Since the normalized XMCD is directly correlated to the magnetic moments there is a similar trend in these values.

However, the ratio of the orbital magnetic moment to the spin magnetic moment is increased for thin layers of Cobalt up to some nm compared to Cobalt bulk. Additionally, the effect is even larger for ultrathin layers of Cobalt up to a few ML so that some structural or electronical relaxation can be assumed. That effect can be correlated to changed interatomic distances compared to bulk as proposed to be caused by possible Co island growth. \cite{lehnert} The resulting total magnetic moments in figure~\ref{fig5} are compared to experimental data for bulk Cobalt \cite{chen1} as well as theoretical calculations of free standing Cobalt (hcp) \cite{lehnert} and are in good agreement with the values found there although different presumptions have to be made for these calculations. Additionally, there are a lot of measurements for Cobalt on metallic substrates or as an intermediate layer on metallic substrates. \cite{vaz2} As these maesurements provide a large range of magnetic moments the comparison is not drawn directly. At lower thicknesses below 9~\r{A} repeated measurements yielded magnetic moments with opposing signs which indicates that there is no remanent magnetization after all. The calculated moments probably result from noise. Given the error bars theses values should be considered to be zero. Altogether, the calculated total magnetic moments at the thicker layers resemble values for the calculated free standing Cobalt better than the one for bulk. 

Data for measurements at 120~K indicate that the onset in magnetism is extrapolated to be at around 9~\r{A}~(5~ML) already. Nevertheless, the full magnetic moment evolves within the next 1.8~\r{A}, which is about the same transition range as obtained on the same spots for room temperature measurements as well (see additional points in figure~\ref{fig5}). This leads to the assumption that the Curie temperature for the thin layers is lower than for thicker layers.

\section{Discussion}

\begin{figure}[!htb]
\begin{center}
\includegraphics[width=8cm]{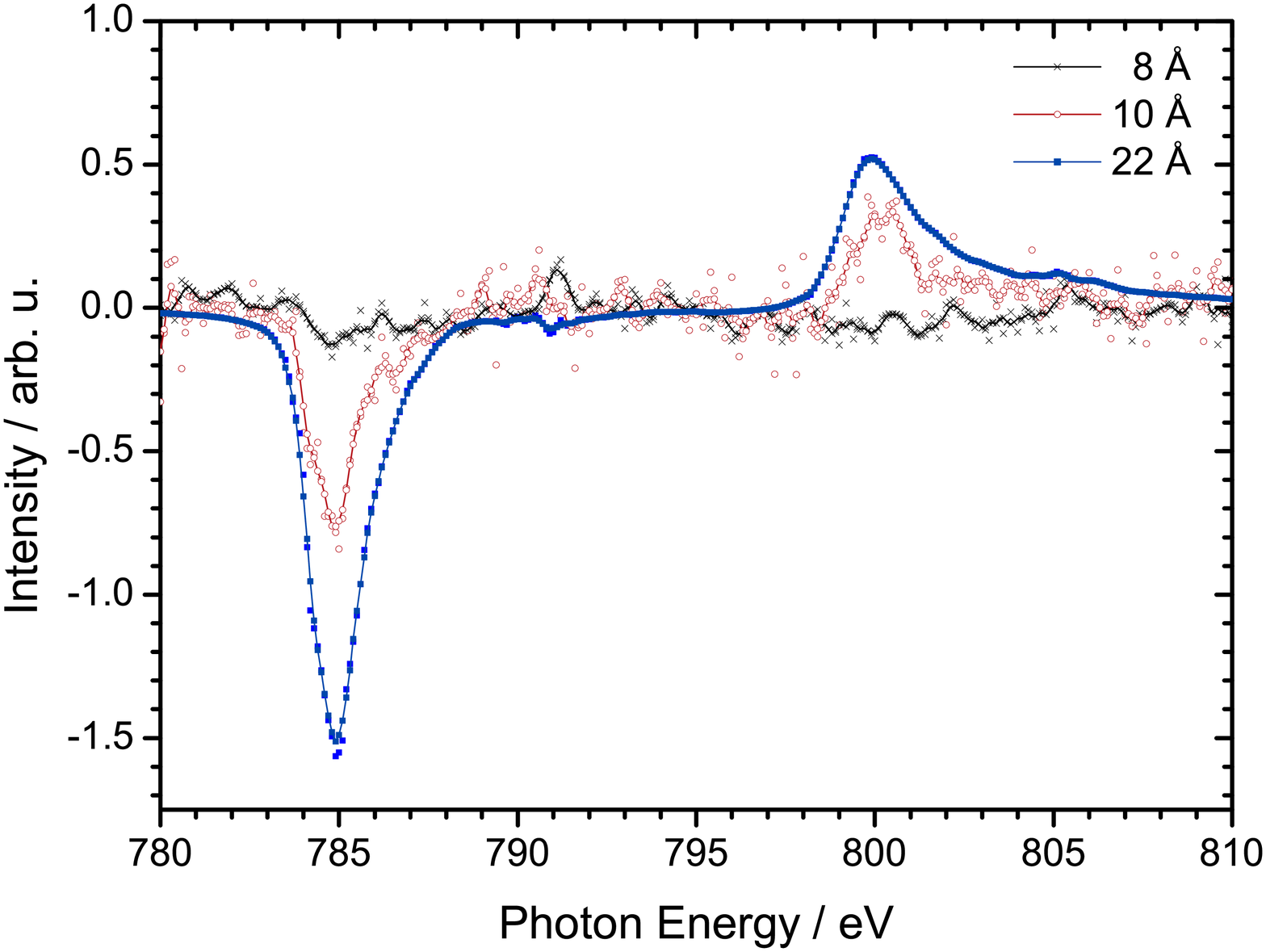}   		 				  
\caption{Thickness dependence of the normalized XMCD spectra measured at 300 K. The spectra show the growing XMCD effect. At 8~\r{A}~/~4.4~ML (black crosses) no XMCD is observed whereas at 10~\r{A}~/~5.5~ML (red open circles) the onset of the difference signal can be seen. With the difference signal at 22~\r{A}~/~12.2~ML (blue closed squares) an example for a thick layer is given. The data points shown are smoothed with a \textsc{Cubic B-Spline} function as a guide to the eye.}
\label{fig4}                                 					 
\end{center}                                 					
\end{figure}

\begin{figure}[!htb]
\begin{center}
\includegraphics[width=8cm]{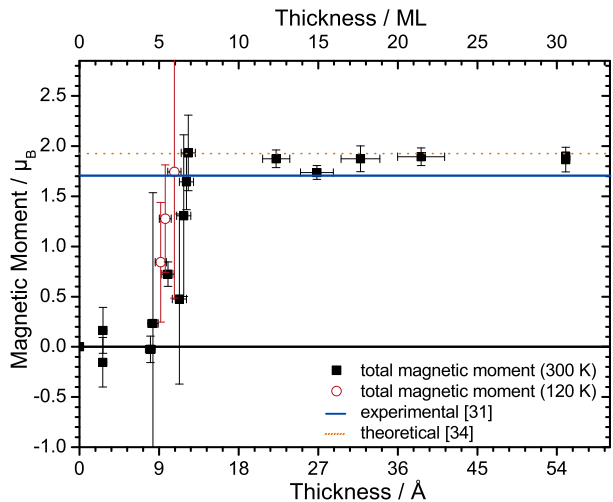} 
\caption{Thickness dependence of the total magnetic moment obtained at 120~K and 300~K. The moments are compared to experimental bulk Cobalt(hcp) data (blue solid line)~\cite{chen1} as well as theoretical calculations for relaxed, free-standing Cobalt(hcp) data (orange dotted line).~\cite{lehnert}}  
\label{fig5}                                 					
\end{center}                                 					
\end{figure}

After thorough data processing the remanent magnetic moments of the Cobalt could be obtained using the sum rule formalism. \cite{IBMsto, jelcsto, jmmmsto} In figures~\ref{fig3}-\ref{fig5} the thickness-dependent behavior of the XMCD, the spin magnetic moments, the orbital magnetic moments, and the total magnetic moments can be seen. There is a clear onset in the XMCD, and hence the magnetic moments, starting at 5~ML of Cobalt. At around 7~ML the full XMCD as well as the full magnetic moment of bulk Cobalt can be observed when an external field of field of 50~mT or 70~mT, respectively, is used for magnetization. Especially for future applications it is important to be able to have working systems at small fields. \textit{In-field} measurements might reveal magnetic moments from the very first layer but are not within the aim of multiferroic coupling devices.

Because of that sharp and abrupt onset magnetically dead layers are to be ruled out. In that case a more gradual development of the calculated magnetic moments with their increasing layer thickness is expected. This is observed for thin Fe or Co layers on V$_2$O$_3$ \cite{sass} or MgO \cite{boubeta} and Co layers on Si(111) or Ge(111) \cite{tsay}. It is more likely that Cobalt grows in small islands with hexagonal configuration which are paramagnetic or even super paramagnetic. It also needs to be mentioned that although the substrate surface was nominally (001), due to the spontaneous polarization the surface usually contains both a and c domains. Ths is expected to influence the growth of Co. Since no very good match of the lattice constants is given anyway, this influence is considered here to be of minor importance. It is probable that islands of Cobalt will form at low thicknesses as this was seen for Fe on Strontiumtitanate upon heating to 800~K. \cite{demund} In fact, since the lattice mismatch to BaTiO$_3$ is larger for Co than this for Fe the effect is expected to appear at lower temperatures. These islands will eventually grow together to form layers plus islands. The main reason for this postulated growth is, thus, the lattice mismatch of Co to barium titanate. This behavior - island growth with resulting super paramagnetism - was observed for Iron on Molybdenum(110) already. \cite{bode} Similar reports for Fe on barium titanate have been given where islands formed on a larger scale. \cite{govind2} Here, the same behavior for Co is assumed. Again it should be pointed out, that for Co this effect is expected to happen at low temperatures as well. If the so formed islands are small and not connected, they exhibit super paramagnetic behavior which does not result in a measurable remanent XMCD signal. \cite{bode} In our understanding the whole layer shows a remanent magnetization once these possible islands coalesce and build up some thicker layers. Thus, the full magnetic moment is observed when a certain thickness is reached. For thin layers of Cobalt on metallic substrates a similar and slightly increased magnetic moment is observed at 9~\r{A} as well. \cite{vaz2} Since the magnetic moments of thin Cobalt layers on metallic substrates cover a range of 1.40~$\mu_B$ to 2.15~$\mu_B$ and there might be specific influence of the oxidic substrate, a direct comparison of these preparations cannot be drawn.

It is likely that the Curie temperature for very thin coalesced islands is lower than room temperature. This assumption is supported by the observation of an earlier onset in remanent magnetism at 120~K which seems to start at a by 0.5 to 1~ML thinner thickness as the one at room temperature (300~K). This is an expected behavior for magnetic materials because of their decreasing Curie temperature with smaller thicknesses. \cite{sriva} Besides this expected effect the change in the onset might be related to the barium titanate phases at low temperatures. In our case at 120~K a rhombohedral crystal lattice with a lattice parameter of a~=~b~=~c~=~4.00~\r{A}. In addition, Iron on barium titanate shows a similar magnetic behavior in XMCD measurements. \cite{govind1}

Added together to total magnetic moments, spin and orbital magnetic moments for thick layers are comparable to known bulk values, with a tendency to be slightly larger than experimentally measured bulk values. \cite{chen1} Nevertheless, the obtained values fit to observations for thin metallic films in general. \cite{vaz2} On the other hand the layers have smaller resulting spin magnetic moments than relaxed, free-standing monolayer of Cobalt whereas the orbital magnetic moments are larger because of the reduced dimensions of such ultrathin layers. \cite{lehnert} This observation would support the postulated island growth as enhanced orbital moments are calculated for strained films or films with increased interatomic distances. \cite{lehnert} In case of island growth such a behavior could be seen. These observations might and should be an effect of the general presence of a substrate as Cobalt on Platinum has an increased orbital moment as well. \cite{lehnert} Altogether, the total magnetic moments rather resemble the values of a free-standing hexagonally closed-packed Cobalt monolayer calculated by theory than those of bulk Cobalt. \cite{chen1, lehnert} It is possible and likely that the Cobalt structure at 55~\r{A}~(30.5~ML) is still not the one of the bulk material leading to yet different total magnetic moments.

\section{Conclusion}

Measuring XMCD spectra at remanent magnetization and room temperature as a function of Co layer thickness shows a clear onset in the XMCD and, hence, the resulting magnetic moment is observed at 10~\r{A}~(5.5~ML). Since that onset is very sharp and abrupt for thicknesses larger than 10~\r{A}, magnetically dead layers are to be ruled out. As pointed out before, it is postulated that there are superparamagnetic islands of Cobalt in the beginning which develop the full magnetic moment once they coalesce. In the future, an STM study might provide insights in this proposed mechanism.

The resulting total magnetic moment is in between known bulk values for Cobalt as well as a relaxed, free-standing Cobalt monolayer. \cite{chen1, lehnert} However, an influence of the substrate cannot be denied. Both, spin and orbital magnetic moments are increased compared to bulk values. On a relative scale the increase in the orbital magnetic moments is more pronounced than the increase in the spin magnetic moments.

Low temperature measurements at 120~K suggest that the onset in magnetism starts for films that are thinner by around 0.5 to 1~ML as shown in figure~\ref{fig5} which might be due to a lower Curie temperature at smaller layer thicknesses.

Evaluation of XAS data of Cobalt on barium titanate below the saturation thickness, where the underlying substrate is not measured anymore, faces some difficulties. Because of the overlap of the Cobalt~\textit{L}~edge and the Barium~\textit{M}~edge the sum rule formalism for the calculation of the magnetic moments of Cobalt cannot be used easily. We have shown a way of careful data processing making it possible to evaluate the magnetic moments of thin films of Cobalt on a substrate including Barium.

These results can be used for future research on the magneto-electric interactions across the interface with ferroelectric polarization studies and in order to prepare and understand the usage of mixed ferrite materials as magnetic part in oxide-only structures (e.g. like in CoFe$_2$O$_4$/BaTiO$_3$ \cite{graefi}). Again, it should be pointed out that remanent magnetization with low magnetic fields is of great importance for future applications so that the work presented here follows the scope of multiferroic devices.

\section*{Acknowledgements}                             
This work was supported by the Sonderforschungsbe-reich SFB 762, 'Functionality of Oxidic Interfaces'. We would like to thank Prof. J. Kirschner for the allocation of the experimental chamber. Many thanks belong to Gunnar \"Ohrwall at MAX-lab in Lund as well as Willy Mahler and Birgit Zada at BESSY II (HZB) for support during the beamtimes. We also acknowledge financial support by the EU (project ELISA I3, MAXlab) and HZB.

\providecommand{\newblock}{}


\begin{thebibliography}{10}
\expandafter\ifx\csname url\endcsname\relax
  \def\url#1{{\tt #1}}\fi
\expandafter\ifx\csname urlprefix\endcsname\relax\def\urlprefix{URL }\fi
\providecommand{\eprint}[2][]{\url{#2}}

\bibitem{ramspal}
Ramesh R and Spaldin N~A 2007 {\em Nature Mater.\/} {\bf 6} 21

\bibitem{duan}
Duan C~G, Jaswal S~S and Tsymbal E~Y 2006 {\em Phys. Rev. Lett.\/} {\bf 97}
  047201

\bibitem{chemo}
Cheong S~W and Mostovoy M 2007 {\em Nature. Mater.\/} {\bf 6} 13

\bibitem{wang1}
Wang L, Maxisch T and Ceder G 2006 {\em Phys. Rev. B\/} {\bf 73} 195107

\bibitem{wang2}
Wang K, Liu J~M and Ren Z 2009 {\em Adv. Phys.\/} {\bf 58} 321

\bibitem{hill2}
Hill N~A 2002 {\em Annu. Rev. Mater. Res.\/} {\bf 32} 1

\bibitem{sterbi}
Sterbinsky G~E, Wessels B~W, Kim J~W, Karaptetrova E, Ryan P~J and Kaevney D~J
  2010 {\em Appl. Phys. Lett.\/} {\bf 96} 092510

\bibitem{brivio1}
Brivio S, Petti D, Bertacco R and Cezar J~C 2011 {\em Appl. Phys. Lett.\/} {\bf
  98} 092505

\bibitem{bocher}
Bocher L, Gloter A, Crassous A, Garcia V, March K, Zobelli A, Valencia S,
  Enouz-Vedrenne S, Moya X, Marthur N~D, Deranlot C, Fusil S, Bouzehouane K,
  Bibes M, Barth\'el\'emy A, Colliex C and S\'ephan O 2012 {\em Nano Lett.\/}
  {\bf 12} 376

\bibitem{graefi}
Gr\"afe J, Welke M, Bern F, Ziese M and Denecke R 2013 {\em J. Magn. Magn.
  Mater.\/} {\bf 339} 84

\bibitem{vaz2}
Vaz C~A~F, Bland J~A~C and Lauhoff G 2008 {\em Rep. Prog. Phys.\/} {\bf 71}
  056501

\bibitem{lawes}
Lawes G and Srinivasan G 2011 {\em J. Phys. D: Appl. Phys.\/} {\bf 44} 243001

\bibitem{govind2}
Govind R, Hari~Babu V, Chiang C~T, Magnano E, Bondino F, Denecke R and
  Schindler K~M 2013 {\em J. Magn. Magn. Mater.\/} {\bf 346} 16

\bibitem{govind1}
Govind R 2014 {\em {Growth, interface and magnetic properties of Fe and
  Fe-oxide ultrathin films on BaTiO$_3$(001) substrates}\/} Ph.D. thesis
  Martin-Luther-Universit\"at Halle-Wittenberg

\bibitem{lahti}
Lahtinen T~H~E, Shirahata Y, Yao L, Franke K~J~A, Venkataiah G, Taniyama T and
  van Dijken S 2012 {\em Appl. Phys. Lett.\/} {\bf 101} 262405

\bibitem{shira}
Shirahata Y, Nozaki T, Venkataiah G, Taniguchi H, Itoh M and Taniyama T 2011
  {\em Appl. Phys. Lett.\/} {\bf 99} 022501

\bibitem{yoo}
Yoo C~S, S\"oderlind P and Cynn H 1998 {\em J. Phys.: Condens. Matter\/} {\bf
  10} L311

\bibitem{carra}
Carra P, Thole B~T, Altarelli M and Wang X 1993 {\em Phys. Rev. Lett.\/} {\bf
  70} 694

\bibitem{sori}
Soriano L, Abbate M, Fern\'andez A, Gonz\'alez-Elipe A~R and Sanz J~M 1997 {\em
  Surf. Interface Anal.\/} {\bf 25} 804

\bibitem{berlich}
Berlich A, Strauss H, Langheinrich C, Chass\'e A and Morgner H 2011 {\em Surf.
  Sci.\/} {\bf 605} 158

\bibitem{pancotti}
Pancotti A, Wang J, Chen P, Tortech L, Teodorescu C~M, Frantzeskakis E and
  Barrett N 2013 {\em Phys. Rev. B\/} {\bf 87} 184116 ISSN 1550-235X

\bibitem{magnu}
Magnuson M, Butorin S~M, Guo J~H and Nordgren J 2002 {\em Phys. Rev. B\/} {\bf
  65} 205106

\bibitem{hudson2}
Hudson L~T, Kurtz R~L, Robey S~W, Temple D and Stockbauer R~L 1993 {\em Phys.
  Rev. B\/} {\bf 47} 10832

\bibitem{lebugle}
Lebugle A, Axelsson A, Nyholm R and M\r{a}rtensson M 1981 {\em Physica
  Scripta\/} {\bf 23} 825

\bibitem{IBMsto}
St\"ohr J and Nakajima R 1998 {\em IBM J. Res. Develop.\/} {\bf 42} 73

\bibitem{jelcsto}
St\"ohr J 1995 {\em J. Electron Spectrosc. and Rel. Phen.\/} {\bf 75} 253

\bibitem{jmmmsto}
St\"ohr J 1999 {\em J. Magn. Magn. Mater.\/} {\bf 200} 470

\bibitem{eal}
{NIST Electron Effective-Attenuation-Length Database} version~1.3
  \urlprefix\url{http://www.nist.gov/srd/nist82.cfm}

\bibitem{imfp}
{NIST Electron Inelastic-Mean-Free-Path Database} version~1.2
  \urlprefix\url{http://www.nist.gov/srd/nist71.cfm}

\bibitem{goer1}
Goering E, Gold S, Bayer A and Schuetz S 2001 {\em J. Synchrotron Rad.\/} {\bf
  8} 434

\bibitem{chen1}
Chen C~T, Idzerda Y~U, Lin H~J, Smith N~V, Meigs G, Chaban E, Ho G~H, Pellegrin
  E and Sette F 1995 {\em Phys. Rev. Lett.\/} {\bf 75} 152

\bibitem{guo}
Guo G~Y, Ebert H, Temmermann W~M and Durham P~J 1994 {\em Phys. Rev. B\/} {\bf
  50} 3861

\bibitem{sass}
Sass B, Buschhorn S, Felsch W, Schmitz D and Imperia P 2006 {\em J. Magn. Magn.
  Mater.\/} {\bf 303} 167

\bibitem{lehnert}
Lehnert A, Dennler S, Blonski P, Rusponi S, Etzkorn M, Moulas G, Bencok P,
  Gambardella P, Brune H and Hafner J 2010 {\em Phys. Rev. B\/} {\bf 82} 094409

\bibitem{boubeta}
Boubeta C~M, Clavero C, Garc\'ia-Mart\'in J~M, Armelles G and Cebollada A 2005
  {\em Phys. Rev. B\/} {\bf 71} 014407

\bibitem{tsay}
Tsay T~S, Yao Y~D, Wang K~C, Cheng W~C and Yang C~S 2002 {\em J. Appl. Phys.\/}
  {\bf 91} 8766

\bibitem{demund}
Demund A 2008 {\em {Interface reactions: Fe/ZnO(000\textit{$\bar{1}$}) and
  Fe/SrTiO$_3$(001)}\/} Ph.D. thesis Universit\"at Leipzig

\bibitem{bode}
Bode M, Kubetzka A, von Bergmann K, Pietzsch O and Wiesendanger R 2005 {\em
  Microsc. Res. Tech.\/} {\bf 66} 117

\bibitem{sriva}
Srivastava P, Wilhelm F, Ney A, Farle M, Haack N, Ceballos C and Baberschke K
  1998 {\em Phys. Rev. B\/} {\bf 58} 5701

\end{thebibliography}
\end{document}